\documentstyle[12pt]{article}
\def\Mdot{ \dot M}

\def\Ms{M_\odot}

\def\rs{\rm s}
\def\rs1{\rm s^{-1}}

\def\rcm{\rm cm}
\def\rcm2{\rm cm^{-2}}
\def\deg{\rm ^{\circ}}
\def\flux{\rerg\ \rcm2\ \rs1}

\def\zeta{A_{Fe}}
\def\dom{\Delta\Omega}

\def\Ms{M_\odot}

\def\rs{\rm s}
\def\rcm{\rm cm}

\def\flux{\rm erg\ cm^{-2}\ s^{-1}}
\def\pflux{\rm  cm^{-2}\ s^{-1}}
\def\csq{\chi^2}
\def\csqn{\chi^2_\nu}

\def\ltsim{\lower 0.5ex\hbox{${}\buildrel<\over\sim{}$}}
\def\gtrsim{\lower 0.5ex\hbox{${}\buildrel>\over\sim{}$}}

\def\etal{{\it et al. }}
\def\apj{{\it Astrophys. J. }}

\def\apjs{{\it Astrophys. J. Suppl. Ser. }}
\def\pasj{{\it Publs. Astron. Soc. Japan }}

\def\aa{{\it Astron. Astrophys. }}
\def\aap{{\it Astron. Astrophys. }}
\def\aas{{\it A\&AS}}
\def\mnras{{\it Mon. Not. R. Astr. Soc. }}

\begin{document}

\baselineskip=24pt

\centerline{\Large \bf Observation of X-ray lines}
\centerline{\Large \bf
from a Gamma-Ray Burst (GRB991216): }
\centerline{\Large \bf 
Evidence
 of Moving Ejecta from the 
Progenitor}

\bigskip
\noindent
{\it L. Piro$^*$, 
G. Garmire $^\dagger$,
M. Garcia$^\ddagger$, 
G. Stratta$^*$, 
E. Costa$^*$,
M. Feroci$^*$,
P.  M\'esz\'aros$^\dagger$,
M. Vietri$^+$,
H. Bradt$^1$,
D. Frail$^2$,
F. Frontera$^3$,
J. Halpern$^4$,
J. Heise$^5$,
K. Hurley$^6$,
N. Kawai$^{7}$,
R. M. Kippen$^8$,
F. Marshall$^9$,
T. Murakami$^{10}$,
V. V. Sokolov$^{11}$,
T. Takeshima$^{12}$,
A. Yoshida$^{7}$
}

\medskip

\noindent
$^*$ Istituto Astrofisica Spaziale, C.N.R., Via Fosso del Cavaliere,
00133 Roma, Italy
$^\dagger$ Department of Astronomy and Astrophysics, 525 Davey Lab, Penn State University, University Park, PA 16802, USA
$^\ddagger$ Center for Astrophysics, Harvard-Smithsonian, 60 Garden St. Cambridge, MA 02138, USA
$^+$  Department of Physics, Universita' Roma Tre,
Via della Vasca Navale 84, 00146 Roma, Italy
$^1$ MIT, 77 Massachusetts Avenue, Cambridge, MA 02139-4307, USA
$^2$ NRAO, 1003 Lopezville Rd., Socorro, NM 87801, USA
$^3$ ITeSRE/C.N.R., Via Gobetti 101, 40129 Bologna, Italy
$^4$ Columbia University, Dept. of Astrophysics, 545 W. 114th St., New York, NY 10027, USA
$^5$ Space Research Organization in the 
Netherlands, Sorbonnelaan 2, 3584 CA, Utrecht, The Netherlands
$^6$ UC Space Science Laboratory, Berkeley, CA 94720-7450, USA
$^{7}$ RIKEN, 2-1 Hirosawa, Wako-Shi, Saitama 351-0198, Japan
$^8$ University of Alabama, NASA/MSFC, Huntsville, AL 35899, USA
$^9$ NASA/GSFC, Code 661, Greenbelt, Maryland 20771, USA
$^{10}$ ISAS, 
3-1-1 Yoshinodai,
Sagamihara, Kanagawa 229-8510, Japan
$^{11}$ SAO/RAS, Special Astrophysical Observatory Nizhnij Arkhyz, Zelenchukskaya, Karachaevo-Cherkesia, Russia 369167 
$^{12}$ NASDA, National Space Development Agency of Japan, WTC Building 2-4-1, Hamamatsu-cho Minato-ku, Tokyo, 105-60 Japan
\medskip

{\bf published in Science, 290, 955 (2000)}

\newpage

{\bf

We report on the discovery  of
two emission features observed in the X-ray spectrum of the afterglow  
of the gamma-ray burst (GRB) of 16 Dec. 1999 by the Chandra X-Ray Observatory.
These features are identified with the Ly$_{\alpha}$ line and 
the narrow recombination continuum
by hydrogenic ions of iron at a redshift $z=1.00\pm0.02$,
providing an unambiguous measurement of the distance
of a GRB.
Line width and intensity 
imply that
the progenitor of the GRB was a massive star system
that ejected, before the GRB event,  $\approx 0.01 \Ms$ of iron  
at a velocity $\approx 0.1 c$, probably by
a supernova explosion.

}

The nature of the  progenitors of GRB's
is an unsettled issue of extreme importance \cite{fryer}.
 The merging of a binary system of
compact objects (such as
black holes, neutron stars, and white dwarfs)
or the collapse of a massive star 
(hypernova or collapsar)
could all deliver the energy required by a GRB,
but observational evidence discriminating
against the various models is still missing. 
This evidence
can be gathered through the measurement of lines
produced by the medium surrounding
the GRB \cite{mr98,bdcl98,pl}. 
However,
while observations of GRB afterglows have 
provided much information on
the  broad band spectral continuum 
and its origin \cite{wg,sp},
they have  not yet given 
results of comparable importance on  spectral lines.
In the optical range current measurements are
inconclusive, because all of the spectral emission
lines observed so far are produced by the host galaxy rather
than at the burst site.
The x-ray range appears more promising because
theoretical computations show that 
only a dense, massive medium close
to the GRB site - such as that expected in the case
of a massive progenitor - could 
produce an iron emission line detectable 
with  current x-ray instrumentation
[7-10].
Indeed, marginal evidence of iron features
has been claimed in two x-ray afterglows
\cite{piro970508,yoshida} but the case is  
still controversial 
not only 
for the limited statistical weight,
but also for the tight upper limits measured in other afterglows
\cite{gb990123asca}, and
for the claimed inconsistency
between the redshift derived for GB970828 in x-rays
and from the host galaxy \cite{gb970828_z},
that could be reconciled only assuming  
different physical conditions in the  two bursts \cite{weth}.
The prospect of gathering unique data on the 
nature of the progenitor
made the search for spectral features one of 
the primary objectives of a Chandra \cite{chandra}
GRB observation program.

The first Chandra observation of a GRB was performed on the
event of 16 Dec. 1999, one of the brightest GRB
ever detected by the Burst And Transient Source Experiment
(BATSE) on board of the Compton Gamma-Ray Observatory, with 
fluence $S_\gamma>
2.5\times 10^{-4}$ erg cm$^{-2}$ above 20 keV
\cite{kippen2}.
 Following the localization of
a strong x-ray afterglow by Rossi X-ray Transient
Explorer 
(guided by the rapid BATSE GRB localization)
and the characterization of its temporal behaviour
\cite{takeshima00},
and a confirming localization by the interplanetary network
\cite{ipn},
 we estimated that the X-ray flux would only decay to
$\approx 10^{-12} \flux$ by the time
Chandra could be re-oriented to point at it. This flux level is high enough
to employ the gratings in conjunction
with the Advanced CCD Imaging Spectrometer
in the Spectroscopic configuration (ACIS-S), and  we selected this
instrument configuration for the observation.
Chandra acquired the
target on 18 Dec., 04:38 U.T., i.e. 37 hours after the GRB, 
and observed it for 3.4 hours. We found a bright x-ray source
\cite{piro991216} 
with a position  (right ascension $(2000)= 05^h09^m31^s.35$,
declination $(2000)= 11\deg17'05''.7$) coincident within the 1.5" error 
with the optical \cite{halpern}
and radio \cite{frail} transients and with a flux consistent
with that expected from the XTE extrapolation.

The spectrum of the x-ray afterglow (Fig. 1) shows
an emission line 
at energy $E=3.49\pm0.06$ keV. 
Due to the ubiquity and prominence of iron lines in
astrophysical objects \cite{piro_line} we argue
that  this line is associated with emission from
iron. Some ambiguity remains in the rest 
energy of the emission.
Iron $K_\alpha$ lines have rest energies ranging from 
6.4 keV (fluorescence of neutral atoms) to 6.7 keV (He-like ions)
or 6.97 keV (H-like ions).  
In those three cases we would obtain redshifts of
$z=0.83\pm0.02$, $z=0.92\pm0.02$ and $z=1.00\pm0.02$, respectively.
In particular,
at higher energies the ACIS-S spectrum shows  evidence
of a recombination edge in emission
at $E=4.4\pm0.5$ keV.
Identifying this feature with the 
iron recombination edge with rest energy of  9.28 keV gives
$z=1.11\pm0.11$, consistent with the highest of the
redshifts  implied by the emission line.

An iron recombination edge at 9.28 keV is indeed expected 
when the iron emission is driven by photoionization and the
medium is heavily ionized by the radiation produced 
by the GRB and its afterglow
\cite{mr98,weth,paerels}. 
If the medium lies in the line of sight, the edge is expected to be seen in
absorption at early times \cite{bdcl98,weth}, and evidence of such a feature
has been  found in another GRB \cite{amati}. 
At later times, when the medium becomes heavily ionized  
and recombination takes place, the edge is seen in emission. This is our case.
In this condition x-ray lines are  produced
almost exclusively through recombination of electrons
on  H-like iron \cite{kallmann82}. 
The  measured intensities of the two features
are also consistent with theoretical expectations
($I_{edge}/I_L\approx 0.93(kT/keV)^{0.2}$ \cite{paerels},
where $I_{edge}$ and $I_L$ are the
intensities of the recombination edge and
emission lines, respectively, and $T$ is the
electron temperature of the gas). 
We therefore conclude that the redshift  of the GRB is
$z=1.00\pm0.02$. We stress that this measurement 
 is  consistent with the most distant   
absorption system (z=1.02) found in the line of
sight towards GRB991216 by optical spectroscopy
\cite{gb991216_z}. 
This system should then be in the host galaxy of the GRB, 
which  
has probably been identified in  deep optical images
\cite{gb991216_host}.

The detection of the line, the measurement
of the distance ($D=4.7\ Gpc$, assuming $H_0=75 km\ s^{-1} Mpc^{-1}$
and $q_0=0.5$)
 and the fact that the driving process is
recombination allow us to derive a lower limit on the mass of
the line emitting medium.
The number of iron atoms $N_{Fe}$ needed to produce the observed
photon line luminosity\footnote{hereafter some quantities are 
expressed as $Y=Y_n\times10^n$} 
 $L=10^{52}L_{52}=8\times 10^{52} 
\ photons\ s^{-1}$ 
is $N_{Fe}=L_L t_{rec}$,
where each of the $N_{Fe}$ iron atom
 produces  $t_{rec}^{-1}$ line photons per second.
The recombination time of iron \cite{mr98}
is $t_{rec}=30 T_7^{1/2}n_{10}^{-1}\ s$,
where $T=10^7 T_7 K$ and $n=10^{10} n_{10} cm^{-3}$ are 
respectively the
temperature and density of the electrons.
The temperature is constrained  
from the width of the recombination edge to be $kT> 1$ keV,
therefore implying $t_{rec}>30 n_{10}^{-1} s$.
The total mass of material in the line 
region can be written 

$M=M_{Fe}/(X_{Fe} 1.8\times10^{-3})>7 X_{Fe}^{-1} L_{52} n_{10}^{-1} \Ms\ $(1).

$X_{Fe}$ is the iron abundance relative to the sun,
where the iron is  a fraction $1.8\times10^{-3}$ of the total mass.   
The requirement that the ionization parameter $\xi$ be high enough to
keep the Fe in a H-like state, i.e. $\xi = 4\pi D^2 F_x/nR^2 > 10^4$
\cite{kallmann82}, allows us to estimate $n_{10}<15$ where we
have used the flux observed by Chandra of $F_X = 2.3 \times 10^{-12}
{\rm erg~cm^{-2}~s^{-1}}$ and assumed that the distance of the gas
from the GRB is $R > 2 \times 10^{15}$cm.
The latter condition derives from the 
presence of the line 1.5 days after the GRB which,
in the cosmological
rest frame of the burst, is $(1+z)^{-1}$
times shorter than observed.
Using this limit on
the  density and the observed line luminosity allows us
to set a lower limit on the mass of: \\

$M \gtrsim 5 X_{Fe}^{-1} \Ms$;$\ M_{Fe}\gtrsim 0.01 \Ms$ \\

This large
mass  is not ejected during the GRB explosion, but 
in an earlier phase. This is  the only possible
condition to have the material
moving at subrelativistic speed (as shown below)
and be illuminated by GRB photons.
The large mass of pre-ejected material
excludes progenitor models
based
on double neutron stars, black hole - neutron star and black hole
- white dwarf.
These systems eject material
long before they actually merge and the progenitors
of these GRB travel far from their formation sites (and their
ejecta) before producing GRB \cite{fryer}.
Conversely, massive progenitors - which evolve
more rapidly - lead to a GRB in a mass-rich environment.

Additional information on the origin of the ejecta
is derived from the line width ($\sigma_L=0.23\pm0.07$ keV, Fig.1).
Thermal broadening 
($\Delta E/E= (kT/A m_p c^2)^{1/2})$ is negligible here, while
Compton broadening  would require a Thompson
optical depth $\tau>1$, which is excluded by \cite{bdcl98}.
The line width
is therefore kinematical, with $v\approx 0.1 c$.
Normal winds from stars are not compatible with
the parameters of the medium.
In fact, the wind density is
$n\approx 10^3 (\Mdot_{-4} R^{-2}_{16} (v/0.1c)^{-1} cm^{-3}$.
Even for a  high value of the mass loss rate
$\Mdot_{-4}= \Mdot/(10^{-4} \Ms yr^{-1})=1$, the density is  orders of
magnitudes lower than that required to produce the observed
line flux.
Weth et al \cite{weth} argued that high density clouds
could be produced from a low-density and low-velocity wind,
but it remains to be assessed how those clouds could
be accelerated to high velocities.
Alternatively,
nearly all the formation scenarios of GRB progenitors
involve
substantial mass loss when the
system is in a common--envelope phase,
a process which is likely to form a disk\footnote{The commom--envelope phase
happens when the hydrogen envelop
of the secondary star expands engulfing
the  compact primary (neutron star or black hole).
Friction and tidal forces cause the compact 
object to spiral in the giant's core,
ejecting the hydrogen envelope preferentially
along the orbital plane, forming a disk \cite{sandquist}}.
Interaction of the  expanding shell of the
GRB with the disk could produce
a shock-heated gas with density and velocity
of the same order of magnitude as needed here
\cite{bf}. The emission from this region
can be represented by pure thermal plasma,
i.e. in thermal and collisional ionization
equilibrium, but in this case
emission from a recombination edge  is negligible
 \cite{bf}.
In fact, to effectively ionize iron atoms,
electrons should have a temperature comparable
to the edge energy, therefore producing
a feature too smeared to be detected.
This may be circumvented if 
the electron population of the
shock-heated plasma does not reach
complete  equilibrium, 
but the conditions under
which this happens have yet to be studied.
The simplest explanation of our results 
is  a mass ejection by the
progenitor with the same velocity implied by the observed line width.
The ejection should have then occurred
 $\approx R/v=$ (i.e., a few months)
before the GRB.

The distribution
of ejecta and the GRB emission are not highly
anisotropic. 
Let  $\dom$ be the solid angle of the medium
illuminated by the GRB and $\Delta R$ its size.
The mass contained in  the line emitting volume
is
$M= \dom R^2 \Delta R n m_p$.
Substituting the limit of the mass derived in the left hand
side of eq(1)
and the limits  $\tau=\Delta R n \sigma_T<1$ and
$\xi>10^4$ we derive

$\dom/4 \pi \gtrsim 60 X_{Fe}^{-1}>0.1$

where the lower limit corresponds to the
extreme case of ejecta of pure iron.
The GRB emission and the distribution of the  
medium around it cannot therefore deviate substantially
from isotropy.
The lower limit we have derived on the
beaming factor \footnote{The beaming factor 
$\Delta\Omega/4/\pi$
is the
fraction of sky over which 
GRB photons are emitted. It is equal to 1 if the
emission is isotropical and $\approx \theta^2/4$
in the case of a jet with opening angle $\theta$} 
is marginally compatible  with the estimate of \cite{halpern}.
We note, however, that our results refer to 
the emission inside the line
emitting region while the measurements
mentioned above are based on the appearence
of a break in the light curve 
2-5 days after the GRB, i.e., when the
fireball  has overcome the line region.
It is then possible that the initial GRB emission is 
isotropic and it is then collimated
by the interaction with a funnel-like medium.
This limit on anisotropy allows  the determination of
the total, isotropic, electromagnetic energy produced by a GRB, which
for GRB991216 is ${\rm E > 7.2 \times 10^{52} erg\ s^{-1}=
 0.04 M_{\odot} c^2}$.
Another important implication of
the previous limit is on 
the iron abundance of the medium, that has 
to be much higher than solar ($X_{Fe}> 60$).
This  high value of the iron abundance
indicates that the ejecta were - at some stage
of the progenitor evolution - produced by 
a supernova explosion  \cite{ww,b}.

In conclusion,  
the most straightforward
scenario that emerges from  all the pieces of evidence 
we have gathered is the following.
A massive
progenitor - like a hypernova or a collapsar \cite{pacz98,w93} -  
ejects, shortly before the GRB, a substantial fraction 
of its mass.
This event is  similar to a supernova explosion,
like in the case of the SupraNova model \cite{vs98}.



\newpage
{\bf Figure 1} The X-ray afterglow
spectrum of GRB991216 
obtained with
the Chandra high-energy gratings (High Energy (HE) and 
Medium Energy (ME)  summed together.
The background is negligible.
The exposure time of the observation was 9700 s.
In order to increase the statistics, 
the grating spectrum has a bin size of 0.25 \AA,
including about 10 resolution elements of the
ME and 20 of the
HE. The dashed line
represents the best fit power law on the
the 0th order ACIS-S spectrum.
The peak (i.e. 2 bins) around 3.5 \AA (E=3.5 keV)
is detected with 4.7 $\sigma$ confidence.
We have verified the robustness of the detection
against the continuum level.
The significance remains above $4\sigma$
even when assuming 
a worst case systematic uncertainty
in the cross-calibration of the two instruments
of 15\% \cite{calib}.
In the inset the region on the line is shown
with a finer binning. The dotted line represents the
best fit continuum model to the  0th 
order ACIS-S spectrum after the addition of a recombination
edge in emission (see Fig.2).
The line parameters (errors on best fit
parameters correspond
to 90\% confidence level for 1 parameter
of interest) are $I_L=(3.2\pm0.8)\ 10^{-5} \pflux$
$E.W.=(0.5\pm.013)$ keV,
width($\sigma_L$)=$(0.23\pm0.07)$ keV, $E=(3.49\pm0.06)$ keV.
The spectrum has been examined
at higher resolution
to confirm the line broadening.
Since each of the spectral bins in the figure
includes several resolution elements of the instrument,
a narrow feature would appear in no more than 1 single bin,
regardless of how fine is the binning, 
while this is not the case.
Deviations around $7\ \AA$  are
$\approx 3 \sigma$, 
and it is 
worth noticing that they are close
to the expected energy (at z=1.0)
of the recombination edge of hydrogen-like
Sulphur. Deviations at $\approx 4.4\ \AA$ are  less than
 3 $\sigma$.

\medskip
{\bf Figure 2}
The X-ray afterglow
spectrum of GRB991216 
obtained with
the Chandra ACIS-S (0th order).
The energy resolution of ACIS-S is 0.1 keV (FWHM)
at 4 keV and the background is negligible on the whole
energy range.
The better response at high energies compared to the gratings
allows us to single out the presence
of a further emission feature. 
Fitting a model (continuous green line)
composed by an
emission edge (blue dashed line) plus  a power law 
(green dashed line) plus line ( orange dashed line) 
provides a satisfactory fit to the data
($\csqn=0.95$, $\nu=26$).
The addition of the edge
improves
the fit by $\Delta\csq/\csqn=16.3$, that
corresponds to a confidence level of 99.5\% (F test).
Best fit parameters of the edge 
are   $E=4.4\pm0.5 $ keV,
$I_{edge}=(3.8\pm2.0)\ 10^{-5} \pflux$
and width $\sigma_{edge}>1 $ keV.
For the power law we derive
($F_X(2-10 keV)= 2.3\ 10^{-12} \flux$, 
photon index $\Gamma=2.2\pm0.2$,
$N_H=(0.35\pm0.15)\ 10^{22}$ cm$^{-2}$, consistent with the 
absorption in our Galaxy ($N_{HG}=0.21\ 10^{22} cm^{-2}$).
Line parameters are consistent with those derived
from the grating. 
Moreover, the edge is  consistent with
the grating data, as shown by the dotted line in the inset of Fig.1.


\begin{thebibliography}{}

\bibitem{fryer}
Fryer C. L., Woosley S. E. \& Hartman D. H. 
\apj, {\bf 526}, 152 (1999). 

\bibitem{mr98}
M\'esz\'aros P. \& Rees M. J.,
\mnras, {\bf 299}, L10.
(1998)

\bibitem{bdcl98}
Boettcher M., Dermer C.D.D., Crider A. W. \& Liang E. D.  
\aap, {\bf 343}, 111, 
(1999).

\bibitem{pl}
Perna R. \& Loeb A., \apj, {\bf 501}, 467 (1998).



  

\bibitem{wg}
Wijers R.A.M.J. \& Galama T.J.,
\apj, {\bf 523}, 177 (1999)

\bibitem{sp}
Sari R., Piran T., \& Narayan R. 
\apj, 497, L17 (1998).


\bibitem{lazzati}
Lazzati D., Ghisellini G. \& Campana S., 
\mnras, {\bf 304}, L31 (1999).

\bibitem{vietri}
Vietri M., Perola G.C., Piro L. \& Stella L., 
{\bf 308}, L29 (2000).



\bibitem{weth}
Weth C., Meszaros P., Kallman T., Rees M., 
\apj
{\bf 534}, 581 (2000).

\bibitem{bf}
Boettcher M. \& Fryer C.L.,
\apj, submitted, preprint available at http://xxx.lanl.gov/abs/astro-ph/
astro-ph/0006076.


\bibitem{piro970508}
Piro L. \etal 
\apj, {\bf 514}, L73 (1999).

\bibitem{yoshida}
Yoshida A. \etal, 
 {\it Proc. of Gamma-ray bursts in the
afterglow era} (eds F. Frontera and L. Piro)
(\aas {\bf 138}, 433 (1999).


\bibitem{gb990123asca}
Daisuke Y. \etal
\pasj, {\bf 52},  (2000). 


\bibitem{gb970828_z}
Kulkarni S. R. \etal 
{\it Proc. of 5th Huntsville GRB Symposium}, in press
(astro-ph/0002168) (2000).





\bibitem{chandra}
Weisskopf, M. C. , O'Dell, S. \& van Speybroeck, L. P. 
{\it Proc of Multilayer and Grazing Incidence X-Ray/EUV Optics III} 
(eds R. B. Hoover and A. B. Walker), 2
(SPIE, 2805, 1996).

\bibitem{kippen2}
Kippen R.M., R.D. Preece, \& T. Giblin,
GCN n.463 \& n. 504 (2000).

\bibitem{takeshima00}
Takeshima T., Markwardt C., Marshall F., Giblin T. \& Kippen R.M.,
GCN n. 478 (2000). 

\bibitem{ipn}
Hurley K. \etal GCN n.484 \& 505 (2000). 


\bibitem{piro991216}
Piro L., Garmire G., Garcia M. \& the CXC team, Marshall F.,
Takeshima T. , GCN n. 500 (2000). 

\bibitem{halpern}
Halpern J.P., \etal 
\apj in press, preprint available at http://xxx.lanl.gov/abs/astro-ph0006206
 (2000).

\bibitem{frail}
Frail D. A., \etal \apj in press,
preprint available at http://xxx.lanl.gov/abs/astro-ph/0003138 (2000). 


\bibitem{piro_line}
Piro L. 
{\it Proc. of UV and X-ray spectroscopy of laboratory and astrophysical plasmas} (eds E. Silver and S. Kahn) 448 (Cambridge Univ. Press, Cambridge 1993).


\bibitem{paerels}
Paerels F., E. Kuulkers, J. Heise, J. in 't Zand, D. Liedahl, 
\apj in press,
 preprint available at http://xxx.lanl.gov/abs/astro-ph/0004188 (2000). 

\bibitem{amati}
Amati L. \etal, Science, this issue.

\bibitem{kallmann82}
Kallmann T. R. \& McCray R. 
\apjs, {\bf 50}, 263 (1983).

\bibitem{gb991216_z}
Vreeswijk P.M. \etal  
GCN n. 496, (2000).

\bibitem{gb991216_host}
Djorgovski S. G. \etal 
GCN n. 510 (2000).


 


\bibitem{sandquist}
Sandquist, E., Taam, R.E., Chen, X., Bodenheimer, P., \& Burkert, A. 
\apj, {\bf 500}, 909 (1998).

\bibitem{ww}
Woosley S. E., \& Weaver T.A. 
\apjs, {\bf 101}, 181 (1995).

\bibitem{b}
Bouchet P. \etal 
\aa,  {\bf 245}, 490 (1991).

\bibitem{pacz98}
Paczynski, B. 
\apj, {\bf 494}, L45 (1998).

\bibitem{w93}
Woosley S. E.
\apj, {\bf 405}, 273 (1993).





\bibitem{vs98}
Vietri M. \& Stella L., \apj, {\bf 507}, L45 (1998).




\bibitem{calib}
See Chandra calibration reports in http://asc.harvard.edu/cal

\bibitem {}
We thank the Chandra team for the support,
in particular F. Nicastro and A. Fruscione
for the help in data reduction. 
We thank for M. Boettcher for his critical
suggestions.
We acknowledge useful
discussions on the observational strategy with G. Ricker
and L.A. Antonelli for comments. MG  acknowledges
the support of NASA Contract NAS8-39073 to the CXC.


\end{thebibliography}
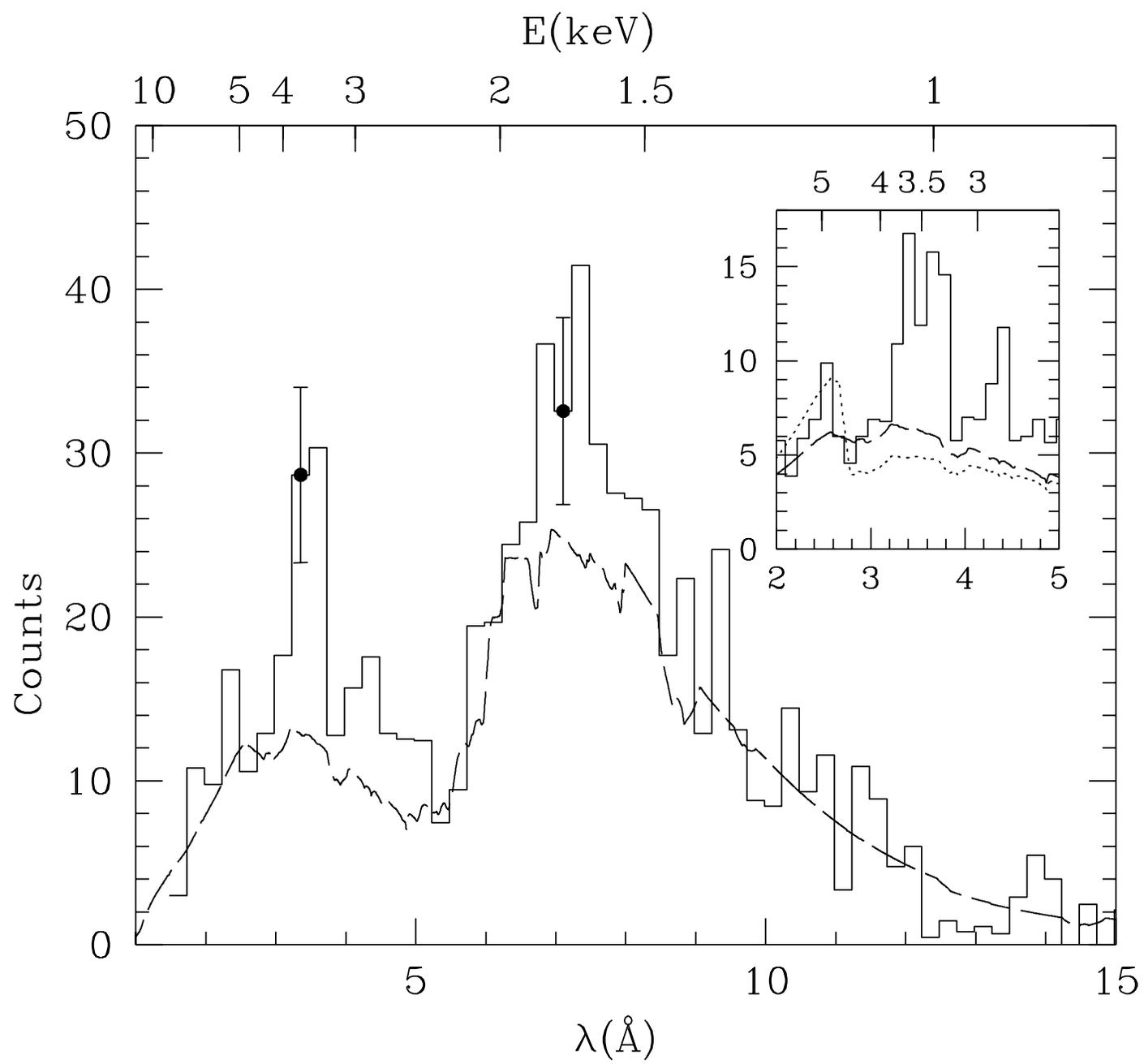
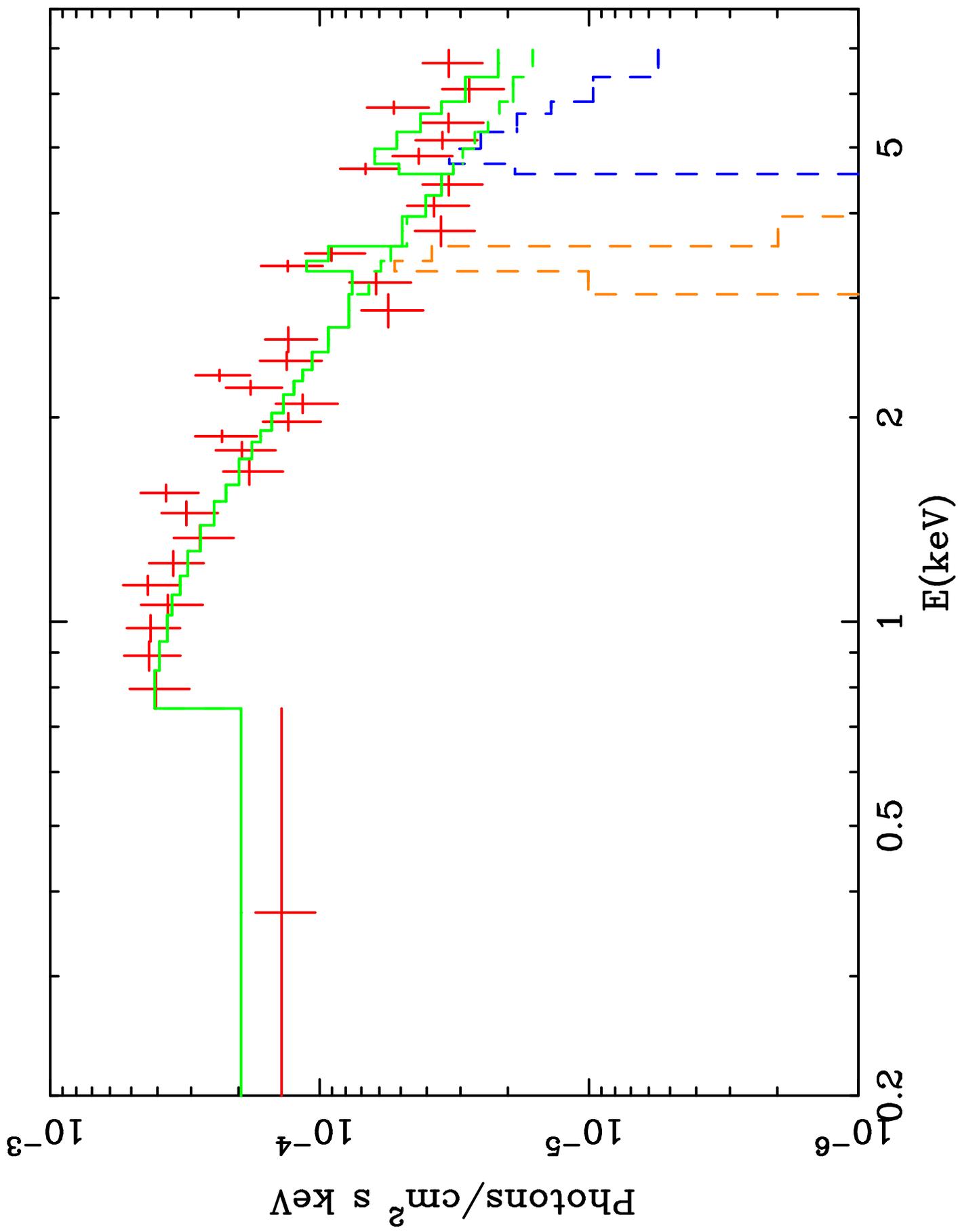
\end{document}